\documentclass[useAMS,usenatbib]{mn2e}
\usepackage{natbib}
\usepackage{amsmath,amsfonts}
\usepackage{epsfig}
\usepackage{mathptmx}
\usepackage{longtable}

\def\reff@jnl#1{{\rm#1\/}}

\def\aj{\reff@jnl{AJ}}                  % Astronomical Journal
\def\araa{\reff@jnl{ARA\&A}}            % Annual Review of Astron and Astrophys
\def\apj{\reff@jnl{ApJ}}                        % Astrophysical Journal
\def\apjl{\reff@jnl{ApJ}}               % Astrophysical Journal, Letters
\def\apjs{\reff@jnl{ApJS}}              % Astrophysical Journal, Supplement
\def\ao{\reff@jnl{Appl.Optics}}         % Applied Optics
\def\apss{\reff@jnl{Ap\&SS}}            % Astrophysics and Space Science
\def\aap{\reff@jnl{A\&A}}                       % Astrophysical Journal
\def\apjl{\reff@jnl{ApJ}}               % Astronomy and Astrophysics
\def\aapr{\reff@jnl{A\&A~Rev.}}         % Astronomy and Astrophysics Reviews
\def\aaps{\reff@jnl{A\&AS}}             % Astronomy and Astrophysics, Supplement
\def\azh{\reff@jnl{AZh}}                        % Astronomicheskii Zhurnal
\def\baas{\reff@jnl{BAAS}}              % Bulletin of the AAS
\def\jrasc{\reff@jnl{JRASC}}            % Journal of the RAS of Canada
\def\memras{\reff@jnl{MmRAS}}           % Memoirs of the RAS
\def\mnras{\reff@jnl{MNRAS}}            % Monthly Notices of the RAS
\def\pra{\reff@jnl{Phys. Rev. A}}         % Physical Review A: General Physics
\def\prb{\reff@jnl{Phys. Rev. B}}         % Physical Review B: Solid State
\def\prc{\reff@jnl{Phys. Rev. C}}         % Physical Review C
\def\prd{\reff@jnl{Phys. Rev. D}}         % Physical Review D
\def\prl{\reff@jnl{Phys. Rev. Lett}}      % Physical Review Letter
\def\pasp{\reff@jnl{PASP}}              % Publications of the ASP
\def\pasj{\reff@jnl{PASJ}}              % Publications of the ASJ
\def\qjras{\reff@jnl{QJRAS}}            % Quarterly Journal of the RAS
\def\skytel{\reff@jnl{S\&T}}            % Sky and Telescope
\def\solphys{\reff@jnl{Solar~Phys.}}    % Solar Physics
\def\sovast{\reff@jnl{Soviet~Ast.}}     % Soviet Astronomy
\def\ssr{\reff@jnl{Space~Sci.Rev.}}     % Space Science Reviews
\def\zap{\reff@jnl{ZAp}}                        % Zeitschrift fuer Astrophysik
\def\nat{\reff@jnl{Nature}}             % Nature

\def\p#1by#2{{\partial{#1} \over \partial{#2}}}
\def\pp#1by#2#3{{\partial^2{#1} \over \partial{#2}\partial{#3}}}
\def\d#1by#2{{{\rm d}{#1} \over {\rm d}{#2}}}
\def\dd#1by#2#3{{{\rm d}^2{#1} \over {\rm d}{#2}{\rm d}{#3}}}

\title[AMI-LA Class I protostars in Taurus]{Radio continuum observations of Class~I protostellar disks in Taurus: constraining the greybody tail at centimetre wavelengths\thanks{We request that any reference to this paper cites ``AMI Consortium: Scaife et~al. 2011''.}}

\label{firstpage}

\author[Scaife et~al.]{
 AMI Consortium: Anna M. M. Scaife$^{1,2}$\thanks{email: a.scaife@soton.ac.uk},
 Jane V. Buckle$^{3,4}$,
 Rachael E. Ainsworth$^2$,
\newauthor
 Matthew Davies$^3$,
 Thomas M. O. Franzen$^{3,5}$,
 Keith J. B. Grainge$^{3,4}$,
 Michael P. Hobson$^3$,
\newauthor
 Natasha Hurley-Walker$^{3,6}$,
 Anthony N. Lasenby$^{3,4}$,
 Malak Olamaie$^3$,
 Yvette C. Perrott$^3$,
\newauthor
 Guy G. Pooley$^3$,
 Tom P. Ray$^2$,
 John S. Richer$^{3,4}$,
 Carmen Rodr{\'i}guez-Gonz{\'a}lvez$^3$,
\newauthor
 Richard D. E. Saunders$^{3,4}$,
 Michel P. Schammel$^3$,
 Paul F. Scott$^3$,
 Timothy Shimwell$^3$,
\newauthor
 David Titterington$^3$,
 Elizabeth Waldram$^3$.
 \vspace{0.03in}\\
$^1$ School of Physics \& Astronomy, University of Southampton, Highfield, Southampton, SO17 1BJ\\
$^2$ Dublin Institute for Advanced Studies, 31 Fitzwilliam Place,
     Dublin 2, Ireland\\
$^3$ Astrophysics Group, Cavendish Laboratory, J J Thomson Avenue,
     Cambridge, CB3 0HE\\
$^4$ Kavli Institute for Cosmology, Cambridge, Madingley Road,
     Cambridge, CB3 0HA\\
$^5$ CSIRO Astronomy \& Space Science, Australia Telescope National Facility, PO Box 76, Epping, NSW 1710, Australia\\
$^6$ International Centre for Radio Astronomy Research, Curtin Institute
of Radio Astronomy, 1 Turner Avenue, Technology Park, Bentley, WA 6845,
Australia \\
}

\date{Accepted ---; received ---; in original form \today}

\pagerange{\pageref{firstpage}--\pageref{lastpage}}

\pubyear{2010}

\begin{document}
%------------------------------------------------------------------------------%
\maketitle

\begin{abstract}
We present deep 1.8\,cm (16\,GHz) radio continuum imaging of seven young stellar objects in the Taurus molecular cloud. These objects have previously been extensively studied in the sub-mm to NIR range and their SEDs modelled to provide reliable physical and geometrical parametres. We use this new data to constrain the properties of the long-wavelength tail of the greybody spectrum, which is expected to be dominated by emission from large dust grains in the protostellar disk. We find spectra consistent with the opacity indices expected for such a population, with an average opacity index of $\beta = 0.26\pm0.22$ indicating grain growth within the disks. We use spectra fitted jointly to radio and sub-mm data to separate the contributions from thermal dust and radio emission at 1.8\,cm and derive disk masses directly from the cm-wave dust contribution. We find that disk masses derived from these flux densities under assumptions consistent with the literature are systematically higher than those calculated from sub-mm data, and meet the criteria for giant planet formation in a number of cases.
\end{abstract}

\begin{keywords}
Radiation mechanisms:general -- ISM:general -- ISM:clouds -- stars:formation
\end{keywords}

\section{Introduction}
\label{sec:intro}

The growth of dust grains from sub-micron to micron sizes in the local environment of protostellar objects is well-probed by observational data in the mid-infrared, however the further growth of these grains to larger sizes is only probed by their sub-mm and cm-wave emission. It is this growth, and the possible coagulation and sedimentation that follows it, which is widely thought to lead to the formation of proto-planetary objects in the dense disks around evolving protostars. Measuring the long wavelength tail of the thermal emission from dust, which is expected to be dominated by such large dust grains, often referred to as \emph{pebbles}, provides an independent measure of the dust composition in the disk through its opacity index ($\beta$), as well as the mass of the accumulated matter in the disk which is an important factor in models of planet formation (e.g. Boss 1998).

Sub-mm measurements towards circumstellar disks are often used to determine their potential for planet formation, as longer wavelength information provides a unique perspective on the cooler dust in the outer part of the disk. It is this region where protoplanets are expected to form. Alternative methods of disk mass estimation, such as spectral line measurements of molecular gas, are complicated by opacity effects (e.g. Beckwith \& Sargent 1993) and require complex models to account for these effects as well as those of depletion (Dutrey 2003; Kamp \& Dullemond 2004). Sub-mm and radio measurements are useful not only as a probe of the disk mass itself, but with multi-wavelength data available, they can also be used to determine the evolution of the opacity index as a function of frequency (Shirley et~al. 2011a;b) and place constraints on the growth of dust grains in such disks. This can be used to determine whether the timescales assumed in current models of planet formation are consistent with observational data. However, to date the disk masses determined from sub-mm data appear to be too low to agree with theoretical models of planet formation (see e.g. Andrews \& Williams 2005), although there are a number of observational caveats associated with these data.

In practice, multiple frequencies are required to constrain the properties of dust emission and draw comparisons with models of radiative transfer (Shirley et~al. 2011a). In the sub-mm the observational properties of the dust derived from its spectral energy distribution (SED) can have contributions from both the envelope and the disk, which need to be separated. In order to probe the properties of the disk alone it is necessary to go to longer wavelengths in order to ensure that it provides the dominant contribution to the SED. However, this transition regime,
between the cm- and sub-mm- wavelengths, is often confused by the presence of multiple emission processes each contributing to the overall SED. For younger protostellar objects, such as Class 0 sources, (partially-) optically thin free-free emission is frequently observed to dominate the cm-wave spectrum; for Class I and II objects the high magnetic field strength on the surface of the central star can lead to gyrosynchrotron emission from quasi-relativistic or thermal plasma. 

In this work we present SEDs for eight Class I protostars in the Taurus molecular cloud. These objects were selected from the sample of Gramajo et~al. (2010) who used multi-frequency data from mid-infrared (MIR) to sub-mm wavelengths to model the SEDs of these objects. We here extend these SEDs to the cm-wave regime with the AMI-LA in order to constrain the long-wavelength properties of the optically thin emission expected to be dominated by their circumstellar disks.

\begin{table*}
\caption{AMI-LA sample of Class I protostars in Taurus. Column [1] source name; column [2] alternative designations for the source; column [3] source Right Ascension; column [4] source Declination; column [5] major axis of the AMI-LA PSF in the combined channel image; column [6] minor axis of the AMI-LA PSF in the combined channel image; column [7] noise on the combined channel image; column [8] date of observation; column [9] primary (flux) calibrator source; \& column [10] secondary (phase) calibrator source. \label{tab:samp}}
\begin{tabular}{lccccccccc}
\hline\hline
Name & Other & RA & Dec & $\theta_{\rm{maj}}$ & $\theta_{\rm{min}}$ & $\sigma_{\rm{rms}}$ & Date & Flux & Phase  \\
     & names & (J2000) & (J2000) & (arcsec) & (arcsec) & ($\frac{\mu{\rm{Jy}}}{\rm beam}$) & dd-mm-yy & Cal. & Cal.  \\
\hline
IRAS04016+2610   & L1489~IRS &    04 04 42.9 &+26 19 02 & 38.9 & 24.6 & 23 & 26-03-11 &	3C147 & J0406+2728 \\
DGTauB           &           &    04 27 02.6 &+26 05 31 & 32.9 & 21.6 & 27 & 07-07-11 &	3C48  & J0429+2724 \\
IRAS04248+2612   & HH~31     &    04 27 57.3 &+26 19 19 & 33.7 & 22.1 & 25 & 28-03-11 &	3C147 & J0429+2724 \\
IRAS04302+2247   &           &    04 33 16.5 &+22 53 21 & 33.5 & 21.5 & 32 & 04-04-11 &	3C147 & J0438+2153 \\
IRAS04325+2402   & L1535~IRS &    04 35 35.4 &+24 08 19 & 33.4 & 23.1 & 22 & 15-04-11 &	3C48  & J0438+2153 \\
IRAS04361+2547   & TMR~1     &    04 39 13.9 &+25 53 21 & 33.6 & 22.2 & 30 & 18-04-11 &	3C48  & J0440+2728 \\
IRAS04368+2557   & L1527     &    04 39 53.6 &+26 03 06 & 32.2 & 22.6 & 30 & 19-04-11 &	3C48  & J0440+2728 \\
\hline
\end{tabular}
\end{table*}

\section{Observations}
\label{sec:obs}

AMI comprises two synthesis arrays, one of ten 3.7\,m
antennas (SA) and one of eight 13\,m antennas (LA),
both sited at the Mullard Radio Astronomy Observatory at Lord's Bridge, Cambridge (AMI Consortium: Zwart
et~al. 2008). The telescope observes in 
the band 13.5--17.9\,GHz with eight 0.75\,GHz bandwidth channels. In practice, the two
lowest frequency channels (1 \& 2) are not generally used due to a lower response in this frequency range and interference from geostationary
satellites. The data in this paper were taken with the AMI Large Array (AMI-LA).

AMI-LA data reduction has been described extensively in previous works (see e.g. AMI Consortium: Zwart et~al. 2009). For this work flux (primary) calibration is performed using short observations of 3C48 and 3C147. From other measurements, we find the flux calibration is accurate to better than
5 ~per~cent (AMI Consortium: Scaife et~al. 2008; AMI Consortium:
Hurley--Walker et~al. 2009). Phase (secondary) calibration is carried out using interleaved observations of
strong point sources, see Table~\ref{tab:samp}. After calibration, the phase is generally stable to
$5^{\circ}$ for channels 4--7, and $10^{\circ}$ for channels 3 and 8. In the following analysis channels $4-8$ have been used. We have applied an increased absolute
calibration error of 10~per~cent to flux densities from channel 8 in order to reflect the poorer phase stability of this channel relative to the others. 

Reduced data were imaged using the AIPS data package, and {\sc{clean}}
deconvolution was performed using the task {\sc{imagr}}. The half power point of the AMI-LA primary beam is $\approx 6$\,arcmin at 16\,GHz, and in what follows we use the convention: $S_{\nu}\propto \nu^{\alpha}$, where $S_{\nu}$ is
flux density, $\nu$ is frequency and $\alpha$ is the spectral index. All errors quoted are 1\,$\sigma$. 

\section{Results}

Flux densities were recovered from both the combined frequency data and the individual frequency channels. The combined frequency image is made in the same manner as the individual channel images, from the weighted combination of multi-channel visibility data. Star-forming regions such as the Taurus molecular cloud often display complex emission structures in the radio and consequently the data were mapped to identify possible confusion with neighbouring sources. In the cases where targets are point-like and isolated (IRAS~04016+2610, 04361+2547, 04302+2247 \& 04325+2402) flux density estimates from the maps were clarified by direct fits to the visibility data; these values were found to deviate by $<5$~per~cent. 

Six of the seven target sources were detected in the continuum data, the exception being IRAS~040248+2612. IRAS~04302+2247 was detected in the continuum data, but at too low significance to recover reliable detections at $>5\,\sigma$ in the individual channels. The remaining five objects were all detected at $>5\,\sigma$ in at least four channels, see Table~\ref{tab:res}.

\begin{table*}
\caption{Measured and derived properties of the AMI-LA Class~I Taurus Sample. Column [1] source name; columns [2-6] measured flux densities at cm-wavelengths; column [7] flux density from the AMI-LA combined channel data; column [8] spectral index across the AMI-LA band. \label{tab:res}}
\begin{tabular}{lcccccccc}
\hline\hline
     &\multicolumn{6}{c}{Freq. [GHz]} & & \\
\cline{2-7}
Name & 8.44$^a$   & 14.62     & 15.37     & 16.12     & 16.87     & 17.63     & $S_{16}$  & $\alpha_{\rm{AMI}}$ \\
     & ($\mu$Jy) & ($\mu$Jy) & ($\mu$Jy) & ($\mu$Jy) & ($\mu$Jy) & ($\mu$Jy) & ($\mu$Jy) &            \\
\hline
IRAS04016+2610  & $343\pm33$ &  $528\pm57$ & $555\pm57$ & $510\pm46$ & $563\pm44$ & $615\pm65$ & $546\pm36$ & $0.65\pm0.27$ \\
DGTauB          & -          &  $523\pm49$ & $555\pm45$ & $595\pm52$ & $645\pm60$ & $735\pm78$ & $647\pm42$ & $1.79\pm0.27$\\ 
IRAS04248+2612  & $<100$     &  -   & -   & -   & -   & -   & $<125$     & - \\ 
IRAS04302+2247  & $160\pm33$ &  -   & -   & -   & -   & -   & $249\pm28$ & - \\ 
IRAS04325+2402  & $<100$     &  $148\pm52$ & $456\pm50$ & $299\pm44$ & $327\pm57$ & $<345$     & $338\pm28$ & $1.49\pm1.10$ \\
IRAS04361+2547  & $140\pm32$ &  $382\pm49$ & $435\pm36$ & $504\pm39$ & -          & $690\pm90$ & $449\pm37$ & $2.82\pm0.72$ \\
IRAS04368+2557  & $810\pm30$$^{b}$ &  $918\pm62$ & $1040\pm62$ & $1194\pm60$ & $1014\pm71$ & $1151\pm130$ & $904\pm54$ & $1.17\pm0.42$ \\ 
\hline
\end{tabular}
\begin{minipage}{0.95\textwidth}
{
$^a$ 8.44\,GHz flux densities from Lucas, Blundell \& Roche (2005).

$^b$ 8.5\,GHz flux density from Melis et~al. (2011).
}
\end{minipage}
\end{table*}

\subsection{Spectral Indices and Dust Disk Masses}

Where possible, power-law spectral indices were fitted to the AMI-LA channel data for each object, see Table~\ref{tab:res}, using the MCMC based  Maximum Likelihood algorithm {\sc metro} (Hobson \& Baldwin 2004). The resulting spectra were all found to be steeply rising with frequency, with a weighted average spectral index $\bar{\alpha}=1.30\pm0.63$. Such steeply rising spectra are compatible with free--free emission, which has a maximum slope of $\alpha=2$ in the optically thick regime. The presence of purely optically thick free--free emission across the AMI-LA band would imply the presence of ultra- or hyper-compact {\sc Hii}, for which a spectral index of $\alpha\approx 1$ is typical. However, both of these are more usually associated with the later stages of more massive star formation. 

Spectral indices with $\alpha>2$, such as that measured for IRAS04361+2546, are consistent with the slope of the SED produced by thermal emission from dust grains. In the Rayleigh-Jeans region the opacity of the thermal emission from dust grains can be well approximated by a power law, $\kappa_{\nu} \propto \nu^{\beta}$. This index $\beta$ is related to the spectral index of flux density measurements, here defined as $S_{\nu} \propto \nu^{\alpha}$, as $\beta \simeq (1+\Delta)\times(\alpha - 2)$, where $\Delta$ is the ratio of optically thick to optically thin emission. This ratio has been previously determined in the region 350\,$\mu$m to 1.3\,mm as $\Delta\simeq 0.2$ (Rodmann et~al. 2006; Lommen et~al. 2007); however, we here assume that for much longer cm-wavelengths $\Delta \rightarrow 0$, and that the emission is entirely optically thin. If the emission at 1.8\,cm is entirely due to thermal dust then the measured $\bar{\alpha}$ would imply values of $\beta\leq0$. Although negative values of $\beta$ are unphysical, values \emph{approaching} zero are proposed for the long wavelength tail of the greybody spectrum from young stellar objects (e.g. Pollack et~al. 1994). Such values are low compared with the canonical value for the interstellar medium of $\beta_{\rm ISM}\approx 1.8$, due to the assumed change in response to the grain size distribution within the dust population, with larger dust grains producing shallower opacity indices. Indeed for grains with a size distribution ${\rm d}n/{\rm d}a\propto a^{-p}$, where the grain size $a\leq a_{\rm max}$ and $a_{\rm max} \geq 3\lambda$, with $\lambda$ the wavelength of observation, the measured opacity index, $\beta(\lambda)\approx(p-3)\beta_{\rm ISM}$. For a typical power-law index of $p=3.5$, grains with $\beta_{\rm ISM}\approx 2$ will have $\beta\approx1$ when $a_{\rm max}> \lambda$; and by extension $\beta\leq1$ when $a_{\rm max}> 3\lambda$ (Draine 2006). Consequently, opacity indices of $\beta\leq1$ indicate the largest grains, which are expected to exist as a consequence of grain growth in protoplanetary disks. Furthermore, power-law indices for grain size distributions as low as $p=3$ have been determined in protoplanetary disks (e.g. Tanaka et~al. 2005) reducing the expected observable value of $\beta$ even further. Indeed, Andrews \& Williams (2005) determined an average $\bar{\beta}=0.06\pm0.02$ in the wavelength range 450--850\,$\mu$m consistent with these predictions, and $\beta<0$ in the range 850\,$\mu$m to 1.3\,mm, the unphysical nature of which they attributed to contamination of the 1.3\,mm data by alternative radio emission mechanisms; a possibility made likely by the spread in protostellar class across their much larger sample. 

\subsection{Spectral Energy Distributions}

For each object in this sample we use data at $12 \leq \lambda\leq 1300\,\mu$m from Gramajo et~al. (2010) and the SED fitting tool of Robitaille et~al. (2007) to fit the sub-mm SED. This tool fits SEDs based on a gridded $\chi^2$ procedure with the option of 200,000 SED models, composed of 20,000 radiative transfer models (Whitney et~al. 2003) viewed at a range of 10 inclination angles. Each SED has contributions from a number of components including the central star itself, the envelope and the disk, the contribution from each of which will vary according to evolutionary stage as indicated by the data.

Since these SEDs are restricted to $\lambda\leq1300\,\mu$m, we make the assumption that the dust SED is dominated at long wavelengths by the contribution from large grains predominantly found in the disk. We therefore extrapolate the disk component of the best-fitting Robitaille et~al. (2007) SED as a power law with a spectral index denoted $\alpha'$ (to distinguish it from the low frequency power law index, $\alpha$) from $\lambda=1300\,\mu$m to the AMI passband to investigate whether the AMI-LA data is consistent with the tail of the greybody spectrum. Under an assumption of optically thin emission at cm-wavelengths, this index is assumed to be related to the opacity index as $\beta' = \alpha'-2$. In summary, data in the centimetre to sub-mm wavelength range are consequently fitted using a double power-law with low frequency data dominated by emission with index $\alpha$ and higher frequency emission dominated by emission with index $\alpha'$. The resulting SEDs are discussed for each object in the following section.

\begin{figure*}
\begin{tabular}{ll}
IRAS~04016+2610: & DG~Tau~B: \\
\includegraphics[angle=0,width=0.45\textwidth]{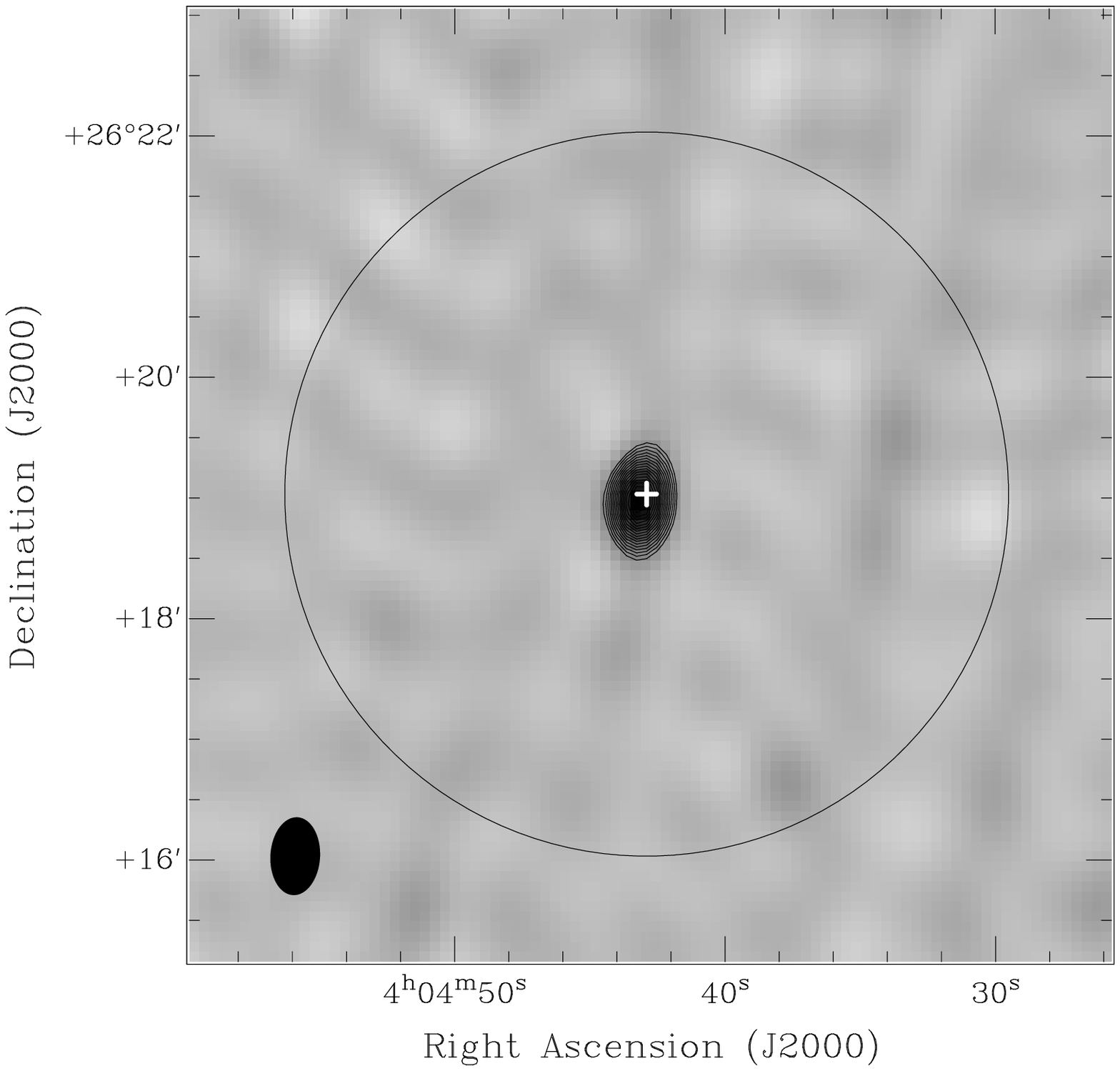} & \includegraphics[angle=0,width=0.45\textwidth]{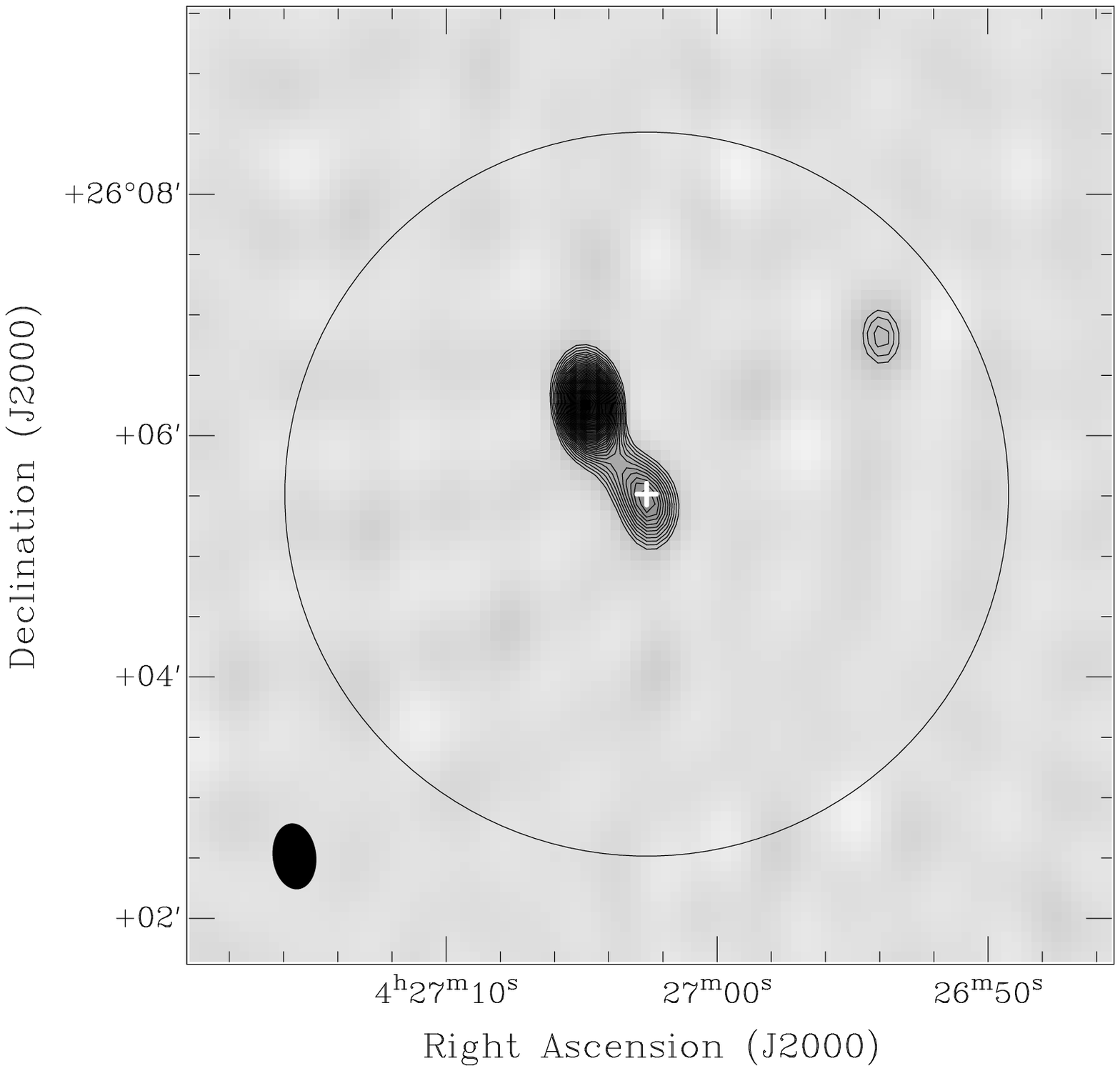}  \\
\end{tabular}
\begin{tabular}{cc}
\includegraphics[angle=-90,width=0.45\textwidth]{./TAU-CI-FIGS/I04016_robspec.ps} & \includegraphics[angle=-90,width=0.45\textwidth]{./TAU-CI-FIGS/DGTauB_robspec.ps}  \\
(a) & (b)\\
\end{tabular}
\caption{\textbf{Top:} AMI-LA 1.8\,cm maps of (a) IRAS~04016+2610 \& (b) DG~Tau~B. Contours are shown in increments of $\sigma$ from 5\,$\sigma$, where $\sigma$ for each object is listed in Table~\ref{tab:samp}. The position of the target object is indicated by a white cross (`+'), the FWHM of the AMI-LA primary beam by a circle, and the synthesized beam is shown as a filled ellipse in the bottom left corner of each map. These maps are uncorrected for primary beam attenuation. \textbf{Bottom:} Spectral energy distributions for (a) IRAS~04016+2610 \& (b) DG~Tau~B. Data points at $\nu >200$\,GHz are taken from Gramajo et~al. (2010), data points at lower frequencies are compiled from the literature (see text for details) and this work, see Table~\ref{tab:res}, with data from the AMI-LA indicated in red in the online version. A best fitting SED from the grid models of Robitaille et~al. (2007) is shown as a solid line at $\nu >200$\,GHz, with the disk component of this model shown as a dashed line. Below $\nu=300$\,GHz this disk component is extrapolated as a power-law, see text for details, and shown as a dot-dash line. A second power-law spectrum is shown fitted to lower frequency radio data at $\nu<10$\,GHz with $\alpha=-1.1$ and $\alpha=-0.1$ for IRAS~04016+2610 and DGTau~B, respectively. The combined spectrum from the two power-law components is shown as a solid line at $\nu <300$\,GHz. \label{fig:sed1}}
\end{figure*}
\begin{figure*}
\begin{tabular}{ll}
 IRAS~04248+2612: & IRAS~04302+2247:\\
 \includegraphics[angle=0,width=0.45\textwidth]{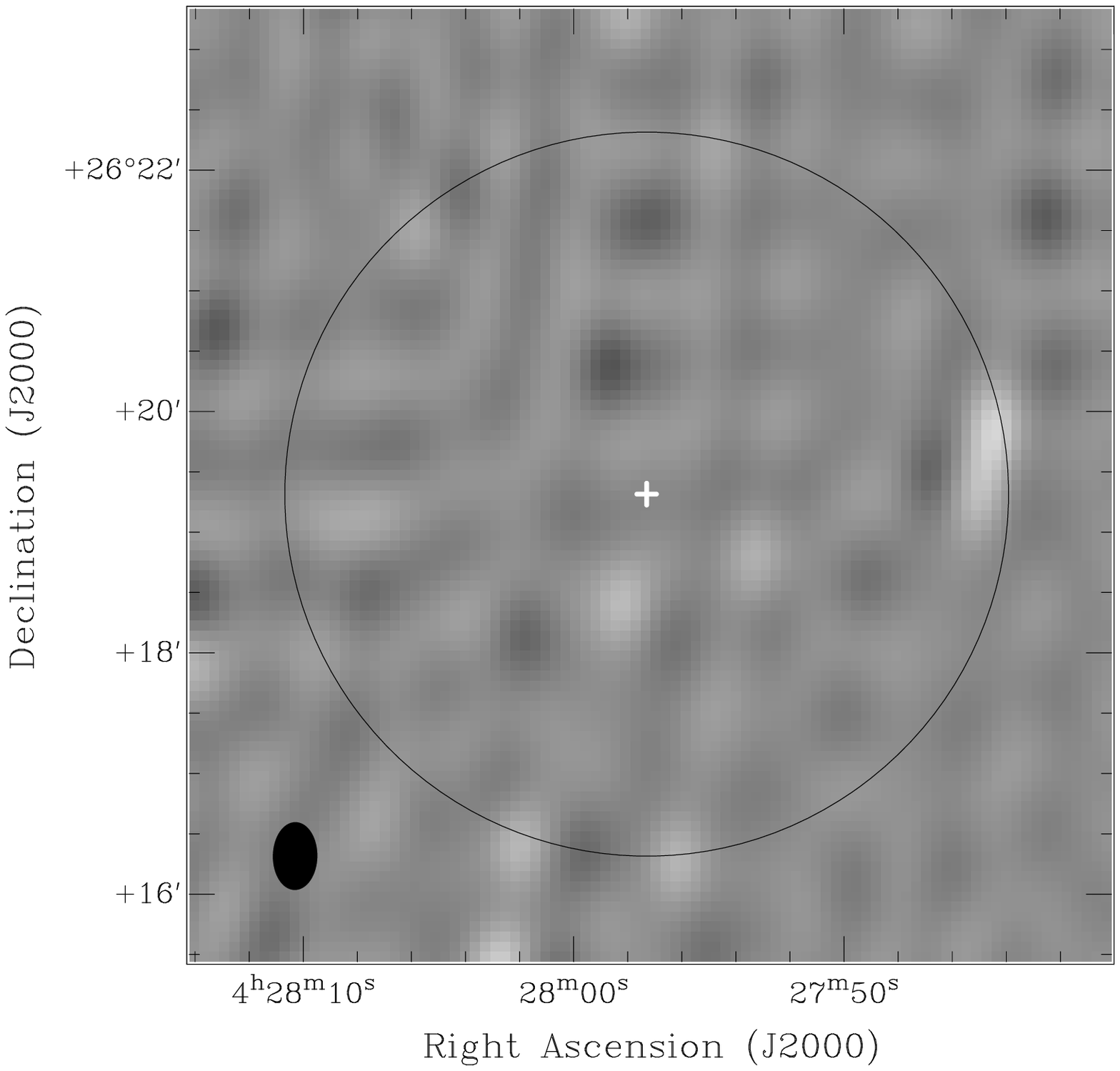} & \includegraphics[angle=0,width=0.45\textwidth]{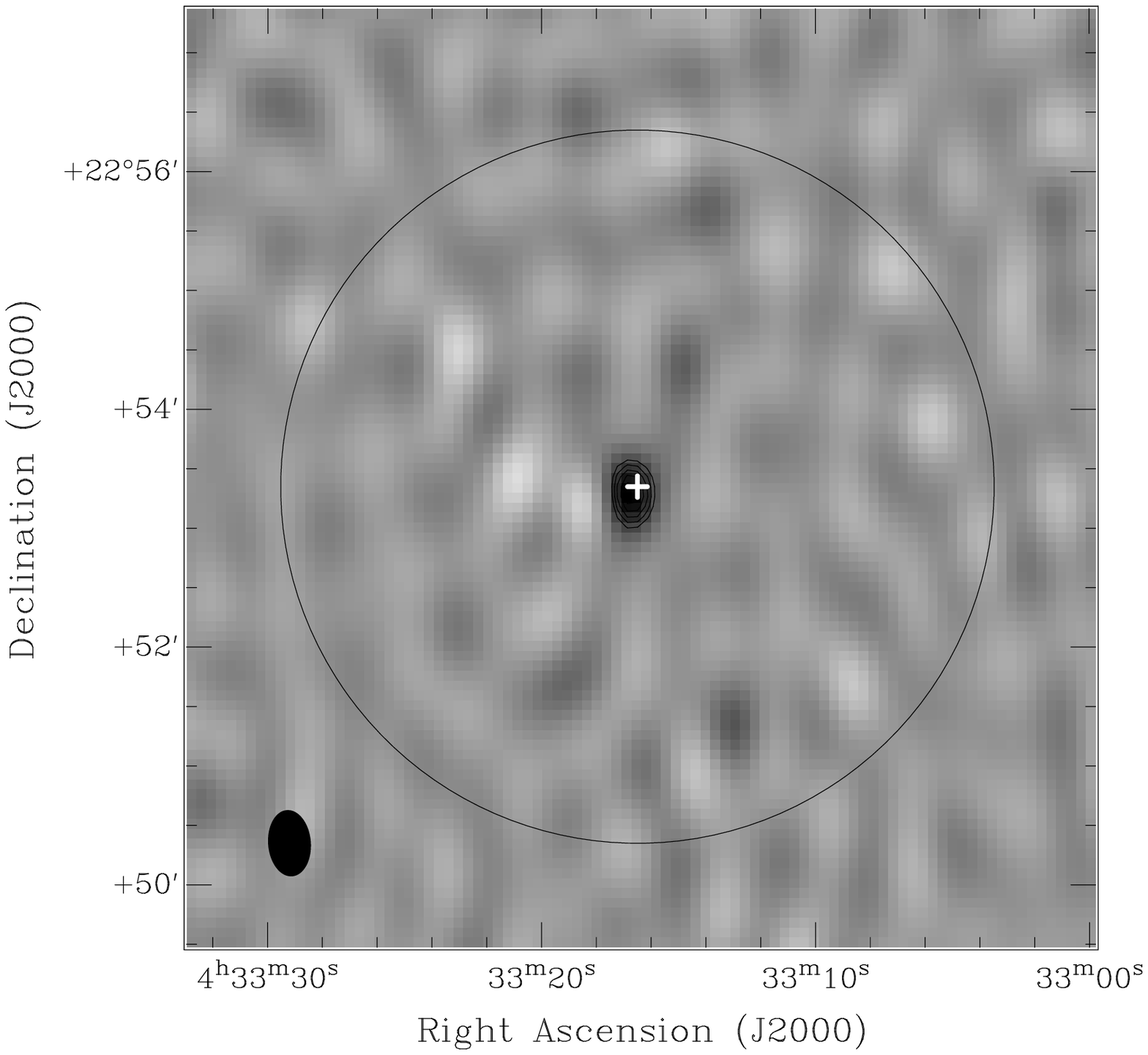} \\
\end{tabular}
\begin{tabular}{cc}
 \includegraphics[angle=-90,width=0.45\textwidth]{./TAU-CI-FIGS/IRAS04248_robspec.ps} & \includegraphics[angle=-90,width=0.45\textwidth]{./TAU-CI-FIGS/IRAS04302_robspec.ps} \\
 (a) & (b) \\
\end{tabular}
\caption{\textbf{Top:} AMI-LA 1.8\,cm maps of (a) IRAS~04248+2612 \& (b) IRAS~04302+2247. Contours are shown in increments of $\sigma$ from 5\,$\sigma$, where $\sigma$ for each object is listed in Table~\ref{tab:samp}. The position of the target object is indicated by a white cross (`+'), the FWHM of the AMI-LA primary beam by a circle, and the synthesized beam is shown as a filled ellipse in the bottom left corner of each map. These maps are uncorrected for primary beam attenuation. \textbf{Bottom:} Spectral energy distributions for (a) IRAS~04248+2612 \& (b) IRAS~04302+2247. Data points at $\nu >200$\,GHz are taken from Gramajo et~al. (2010), data points at lower frequencies are compiled from the literature (see text for details) and this work, see Table~\ref{tab:res}, with data from the AMI-LA indicated in red in the online version. A best fitting SED from the grid models of Robitaille et~al. (2007) is shown as a solid line at $\nu >200$\,GHz, with the disk component of this model shown as a dashed line. Below $\nu=300$\,GHz this disk component is extrapolated as a power-law, see text for details, and shown as a dot-dash line. A second power-law spectrum is shown fitted to lower frequency radio data at $\nu<10$\,GHz with $\alpha=-0.1$ for IRAS~04302+2247. The combined spectrum from the two power-law components is shown as a solid line at $\nu <300$\,GHz. \label{fig:sed2}}
\end{figure*}

\subsubsection{IRAS~04016+2610} Lucas, Blundell \& Roche (2000; hereinafter LBR2000) previously detected IRAS~04016+2610 at 3.5 and 2\,cm, and placed a 3\,$\sigma$ upper limit on the 6\,cm flux density from IRAS~04016+2610 of $S_{4.89\,\rm{GHz}}<260\,\mu$Jy. However, they qualified this non-detection as an effect of the high spatial resolution of their MERLIN observations as demonstrated by a lower spatial resolution detection by Rodr{\'i}guez et~al. (1989) with $S_{4.89\,\rm{GHz}}=500\pm50\,\mu$Jy. The measured flux densities from the detections at these frequencies imply a radio spectrum with $\alpha_{4.89}^{8.44}=-0.69\pm0.70$, suggesting a non-thermal emission mechanism. The 2\,cm flux density of $S_{15\,\rm{GHz}}=700\pm100\,\mu$Jy from Rodr{\'i}guez et~al. (1989) is broadly consistent with that measured by the AMI-LA, and that of LBR2000: $S_{15\,\rm{GHz}}=520\pm40\,\mu$Jy. The wider cm-wave spectrum incorporating all these data, see Fig.~\ref{fig:sed1}(a), may be fitted assuming a dominant contribution to the SED with $\alpha<-1.1$ at $\nu<10$\,GHz, above which frequency the greybody tail from thermal dust emission becomes dominant. We find a best fitting extrapolation to the disk SED with $\alpha'=2.4~(\beta'=0.4)$, see Fig.~\ref{fig:sed1}(a). At face value, these spectral indices imply an SED where a non-thermal emission mechanism is dominant at low radio frequencies. 

Synchrotron emission would be unusual in the vicinity of protostellar objects as it requires both high magnetic field strengths and a population of ultra-relativistic particles. A spectral index of $\alpha=-1.1$ is broadly consistent with gyrosynchrotron emission from a quasi-relativistic power-law population of electron plasma, which is expected to exhibit steeply falling spectral indices (Dulk 1985). Gyrosynchrotron has been invoked as a potential radio emission mechanism for a number of protostellar objects (e.g. Ray et~al. 1996). The position of the peak in the spectrum of such emission is a strong function of the magnetic field strength. A spectrum which peaks below 10\,GHz, such as that of IRAS~04016+2610, is likely to arise from a magnetic field with a magnitude $\sim10$\,G. The magnetic field strength on the surface of Class I objects is thought to be of order $10^3$\,G, as seen towards the Class I object WL~17 (Johns-Krull et~al. 2009); a field strength two orders of magnitude smaller is expected from the inner radius of the circumstellar disk (Donati \& Landstreet 2009) and this may be the source of the gyrosynchrotron emission suggested here.

\begin{figure}
\centerline{\includegraphics[angle=-90,width=0.45\textwidth]{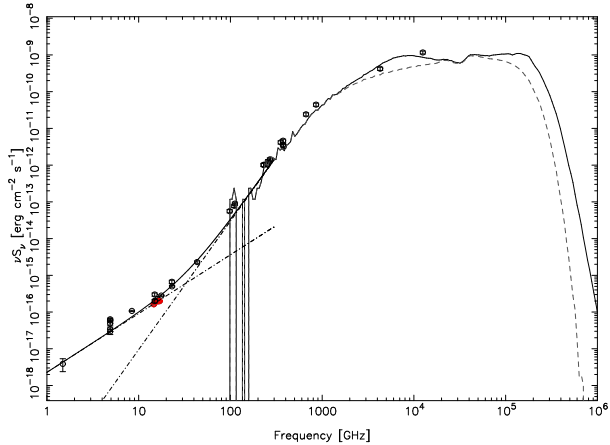}}
\caption{Spectral energy distribution of DG~Tau~A. Millimetre and sub-millimetre data are taken from the literature, see text for details, archival centimetre radio data from Wendker et~al. (1995) are shown, with data points from the AMI-LA (this work) indicated in red in the online version. A best fitting model from the Robitaille et~al. (2007) grid is shown as a solid line with the disk component indicated as a dashed line. Below 300\,GHz this disk component extrapolated as a power-law with $\alpha'=2.5 (\beta'=0.5)$ and a free--free component with $\alpha=0.6$ are shown as dot-dash lines. The combined spectrum from the two power-law components is shown as a solid line. \label{fig:dgtaua}}
\end{figure}

\subsubsection{DG~Tau~B} A variable free--free component is suggested by the SED of DG~Tau~B, see Fig.~\ref{fig:sed1}(b). The magnitude of the emission measured with the AMI-LA is only $\approx$70\,~per~cent of that measured previously by Rodr{\'i}guez et~al. (1996), who found $S_{15\,\rm{GHz}}=0.8\pm0.1$\,mJy for this source, see Fig.~\ref{fig:sed1}(b), although the data are consistent within 3\,$\sigma$. Variability is also evident in the spectrum of the nearby source DG~Tau~A, which lies immediately to the north-east of DG~Tau~B. The spectrum for this object is presented in Fig.~\ref{fig:dgtaua}, where the spread in measurements at 5\,GHz illustrates the degree of variability.

The sub-mm to millimetre SED of DG~Tau~B is poorly constrained due to its proximity to DG~Tau~A, however we find that the spectrum at frequencies $\nu>20$\,GHz can be well described by a power-law with $\alpha=2.0 (\beta\approx0.0)$. This would imply a disk component with emission approximately one order of magnitude higher than that of the best-fitting grid SED found by Gramajo et~al. (2010) and that the disk contribution is significant in DG~Tau~B. We illustrate this scaled disk model as a dashed line in Fig.~\ref{fig:sed1}(b).

Unlike the Class I sources targeted by this sample DG~Tau~A can be seen to possess a more shallowly rising spectrum in the cm to mm-wave regime. This spectrum is nominally well fitted by an index of $\alpha=0.6$, see Fig.~\ref{fig:dgtaua}, the canonical value for partially optically thick free--free emission from a spherical stellar wind (Panagia \& Felli 1975; Reynolds 1986). DG~Tau~A is a Class~II source with a well known wide-tailed molecular outflow, thought to host a large disk wind (Agra-Agoade et~al. 2009), which may be providing the source of this emission. Radio flux densities for this source are taken from the catalogue of Wendker (1995). We do not include DGTau~A in our wider analysis as it is not a member of the original sample.

\subsubsection{IRAS~04248+2612} IRAS~04248+2612 is a non-detection at 1.8\,cm, which constrains an extrapolation of the greybody spectrum to having a spectral index $\alpha'>2.7~(\beta'>0.7)$, see Fig.~\ref{fig:sed2}(a).

\subsubsection{IRAS~04302+2247} IRAS~04302+2247 is detected at $>5\,\sigma$ in the AMI-LA continuum data only and it is therefore not possible to measure a spectral index at cm-wavelengths directly from the AMI-LA data. Previously it was also detected at 8.44\,GHz by LBR2000, who did not detect it at 2 or 6\,cm. Assuming that the detection at 8.44\,GHz represents the contribution of an optically thin free--free emission component we fit the spectrum shown in Fig.~\ref{fig:sed2}(b). Assuming a canonical optically thin spectral index of $\alpha=-0.1$ for the free--free component, this model is compatible with the AMI-LA continuum data point assuming an extrapolation of the best fitting disk component to the SED with a spectral index $\alpha'=2.3~(\beta'=0.3)$. However we note that the low frequency spectral parametres for this object remain largely unconstrained by these data.

\subsubsection{IRAS~04325+2402} The AMI-LA image of IRAS~04325+2402 is dynamic range limited by the bright radio source at the half power point of the primary beam in the north-east of this field, see Fig.~\ref{fig:sed3}(a). We tentatively associate this source with the emission line star Coku-Tau~3. As a result the data, and hence image quality per channel, for this source are comparatively poorer than remainder of the sample and there is increased scatter in the cm-wave spectrum. This is reflected in a much larger uncertainty on the cm-wave spectral index. We find that the cm-wave flux density can be accounted for entirely by an extrapolation of the best-fitting disk component from the Robitaille et~al. (2007) model with $\alpha'=2.1$ ($\beta'=0.1$). This conclusion is also supported by a non-detection of this object at 8.44\,GHz by LRB2000. 

\subsubsection{IRAS~04361+2547} IRAS~04361+2547, see Fig.~\ref{fig:sed2}©, is found to have a steeply rising spectrum across the AMI band with $\alpha_{\rm{AMI}}=2.82\pm0.72$. Such a spectrum is too steep to be attributed to free-free emission, however it is similar to that found for IRAS~16293-2422B (Rodr{\'i}guez et~al. 2005)

\begin{figure*}
\begin{tabular}{ll}
 IRAS~04325+2402: & IRAS~04361+2547: \\
  \includegraphics[angle=0,width=0.45\textwidth]{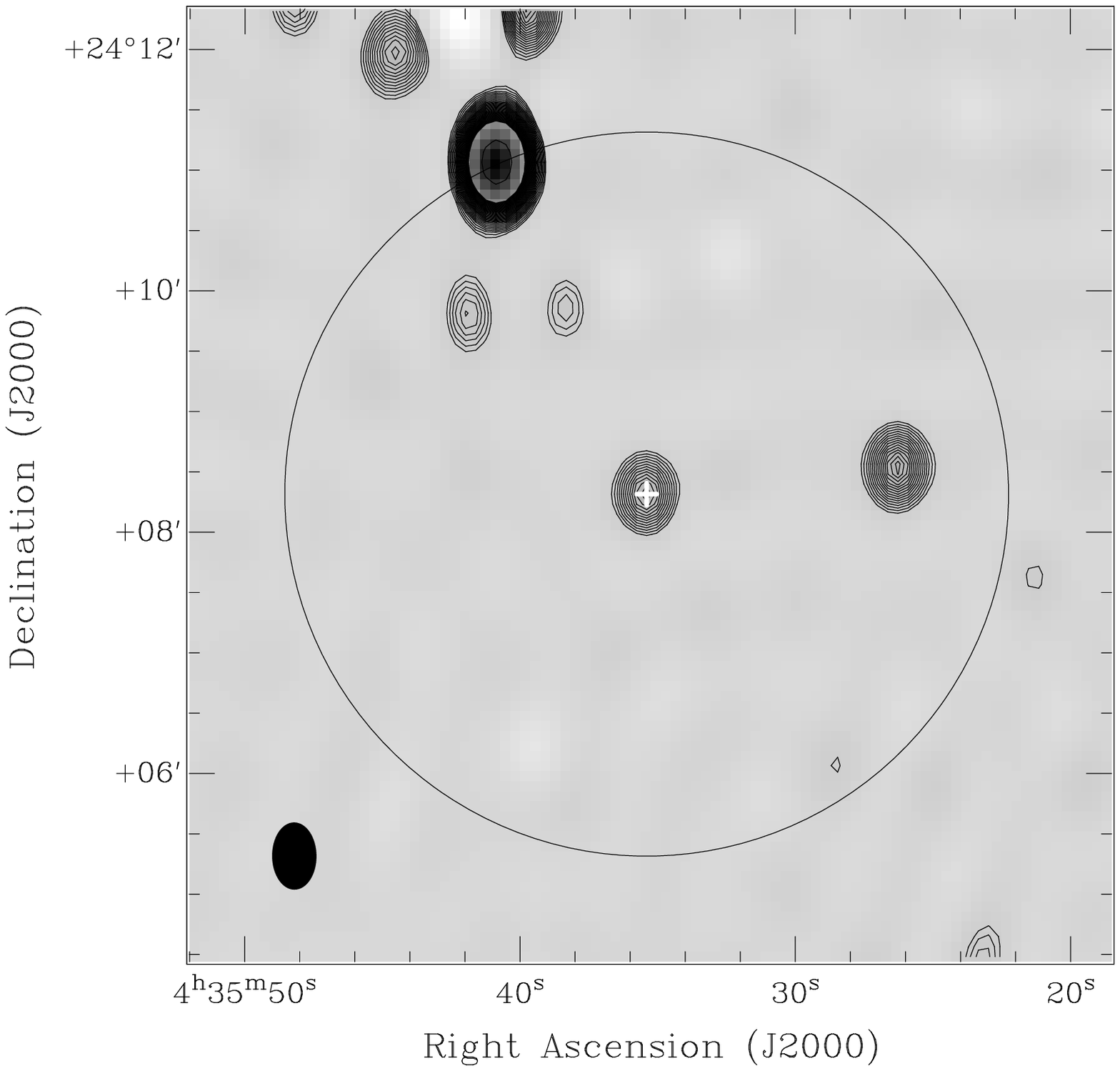} & \includegraphics[angle=0,width=0.45\textwidth]{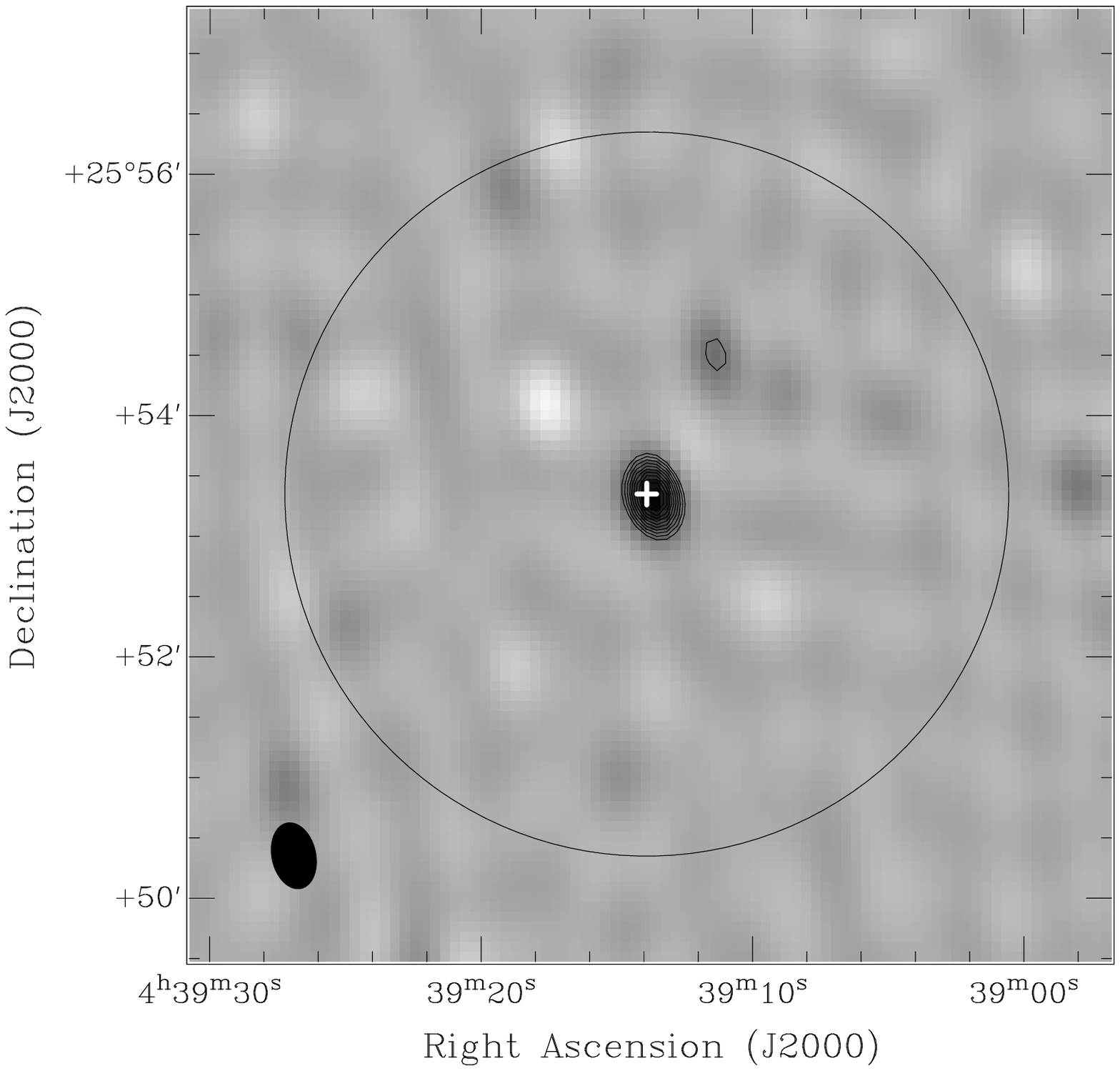} \\
\end{tabular}
\begin{tabular}{cc}
  \includegraphics[angle=-90,width=0.45\textwidth]{./TAU-CI-FIGS/IRAS04325_robspec.ps} & \includegraphics[angle=-90,width=0.45\textwidth]{./TAU-CI-FIGS/IRAS04361_robspec.ps} \\
(a) & (b) \\
\end{tabular}
\caption{\textbf{Top:} AMI-LA 1.8\,cm maps of (a) IRAS~04325+2402b \& (b) IRAS~04361+2547. Contours are shown in increments of $\sigma$ from 5\,$\sigma$, where $\sigma$ for each object is listed in Table~\ref{tab:samp}. The position of the target object is indicated by a white cross (`+'), the FWHM of the AMI-LA primary beam by a circle, and the synthesized beam is shown as a filled ellipse in the bottom left corner of each map. These maps are uncorrected for primary beam attenuation. \textbf{Bottom:} Spectral energy distributions for (a) IRAS~04325+2402b \& (b) IRAS~04361+2547. Data points at $\nu >200$\,GHz are taken from Gramajo et~al. (2010), data points at lower frequencies are compiled from the literature (see text for details) and this work, see Table~\ref{tab:res}, with data from the AMI-LA indicated in red in the online version. A best fitting SED from the grid models of Robitaille et~al. (2007) is shown as a solid line at $\nu >200$\,GHz, with the disk component of this model shown as a dashed line. Below $\nu=300$\,GHz this disk component is extrapolated as a power-law, see text for details, and shown as a dot-dash line. In the case of IRAS~04361+2547 a second power-law spectrum is also shown fitted to lower frequency radio data at $\nu<10$\,GHz with $\alpha=-3.5$. The combined spectrum from the two power-law components is shown as a solid line at $\nu <300$\,GHz. \label{fig:sed3}}
\end{figure*}

IRAS~04361+2547 was previously detected in the radio at 4.89\,GHz (Terebey et~al. 1992) and at 8.44\,GHz by LBR2000 who also made a tentative detection at 2\,cm. With a measured flux density of $S_{4.89\,\rm{GHz}}=230\pm36\,\mu$Jy (Terebey et~al. 1992), a steeply falling spectral index is found below 10\,GHz with $\alpha_{4.89}^{8.44}=-0.92\pm0.47$. Such a spectrum implies a non-thermal emission process such as synchrotron emission, however this would be unusual for a protostellar object. Indeed, when considered as part of the wider cm-wave SED, for this emission to be compatible with that seen at 16\,GHz by the AMI-LA it must possess a spectrum with $\alpha<-3.5$, see Fig.~\ref{fig:sed3}(b). As described previously, such a spectral index would be typical of gryosynchrotron emission from a power-law population of electrons; it is also not incompatible with that of thermal gyrosynchrotron emission, which can have an even more steeply inverted spectrum with $\alpha_{\rm{TH}}=-8$. 

The inference of a non-thermal emission mechanism from the SED does not take account of any possible variable free--free emission from these objects. The observations at 4.89 and 8.44\,GHz were taken 7 years apart, and over a decade prior to those of the AMI-LA. Variable radio emission has been observed from a variety of protostellar objects (e.g. Shirley et~al. 2006) and it is not certain that spectral behaviour determined from non-simultaneous measurements in this region of the spectrum can be relied upon. We note that the tentative detection at 2\,cm of IRAS~04361+2547 by LBR2000 is broadly consistent with the AMI-LA flux densities (albeit at low significance, $<4.5\,\sigma$), and was made at a similar epoch to that at 8.44\,GHz which can also be accounted for by an extrapolation of the AMI-LA power-law. This may imply that the spectral behaviour of IRAS~04361+2547 was similar at the LBR2000 epoch to that of the AMI-LA, but different at the epoch of the 4.89\,GHz observations. Data at frequencies $>5$\,GHz can be accounted for by an extrapolation of the greybody tail using a power-law with $\alpha'=2.0 (\beta'=0.0)$, and this is illustrated in Fig.~\ref{fig:sed3}(b).

\subsubsection{IRAS~04368+2557} 
\begin{figure}
\centerline{IRAS~04368+2557:}

\centerline{\includegraphics[angle=0,width=0.45\textwidth]{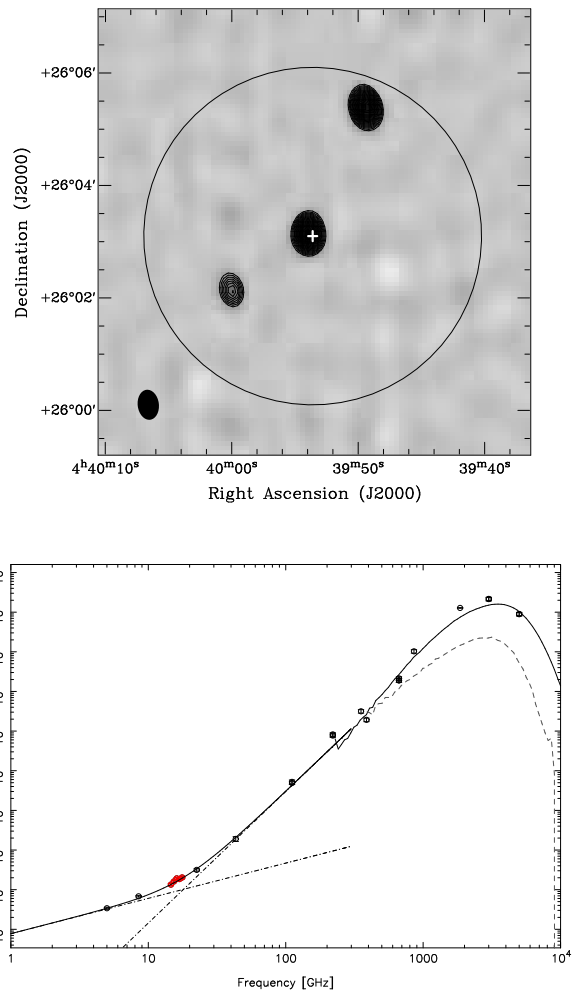}}

\centerline{\includegraphics[angle=-90,width=0.45\textwidth]{./TAU-CI-FIGS/IRAS04368_robspec.ps}}
\caption{\textbf{Top:} AMI-LA 1.8\,cm map of IRAS~04368+2557. Contours are shown in increments of $\sigma$ from 5\,$\sigma$, where $\sigma$ for is listed in Table~\ref{tab:samp}. The position of the target object is indicated by a white cross (`+'), the FWHM of the AMI-LA primary beam by a circle, and the synthesized beam is shown as a filled ellipse in the bottom left corner of the map. This map is uncorrected for primary beam attenuation. \textbf{Bottom:} Spectral energy distribution for IRAS~04368+2557. Data points at $\nu >200$\,GHz are taken from Gramajo et~al. (2010), data points at lower frequencies are compiled from the literature (see text for details) and this work, see Table~\ref{tab:res}. A best fitting SED from the grid models of Robitaille et~al. (2007) is shown as a solid line at $\nu >200$\,GHz, with the disk component of this model shown as a dashed line. Below $\nu=300$\,GHz this disk component is extrapolated as a power-law, see text for details, and shown as a dot-dash line. A power-law radio spectrum is also shown fitted to lower frequency radio data at $\nu<10$\,GHz with $\alpha=-0.11$, see text for details. \label{fig:i04368}}
\end{figure}

IRAS~04368+2557 is defined by Gramajo et~al. (2010) as a borderline Class~0/I object and this is evident from the enhanced envelope contribution to its SED compared with the other sources in this sample, see Fig.~\ref{fig:i04368}. This object is also part of the EVLA survey of edge-on protoplanetary disks (Melis et~al. 2011), which provides flux density measurements for this source at 5, 8.5, 22.5 and 43.3\,GHz. These data are combined with the new AMI-LA measurements presented here and are shown in Fig.~\ref{fig:i04368}. Melis et~al. (2011) fitted data at $\nu\leq 8.5$\,GHz and $>8.5$\,GHz separately and constrained the low frequency radio emission to have a spectral index of $\alpha_5^{8.5}=0.33\pm0.17$ and the high frequency emission to have an index of $\alpha_{\nu>8.5}=2.87\pm0.17$. In this work we fit data from 5\,GHz to 350\,GHz jointly using the sum of two power-laws. There is a degeneracy between the fitted indices and, using MCMC sampling to explore this degeneracy, we find that the maximum likelihood spectral indices are $\alpha=-0.11\pm0.09$ and $\alpha'=2.32\pm0.04$, where (as previously in this work) $\alpha$ and $\alpha'$ represent the spectral indices of power-laws which dominate at low frequency and high frequency, respectively. A spectral index of $\alpha=-0.11$ is typical of optically thin free--free emission. Melis et~al. (2011) define IRAS~04368+2557 as a Class~0 object and a free--free component would not be unusual for Class~0 objects which predominantly have radio counterparts. We find an opacity index $\beta'\approx\alpha'-2=0.32\pm0.04$ substantially shallower than that of Melis
et al. (2011), and consistent with the presence of large dust grains, or
pebbles, in the disk.

\begin{table*}
\caption{Opacity indices and disk masses for the sample. Column [1] lists the target name; column [2] the spectral index measured from AMI-LA data alone; column [3] the spectral index dominant at low frequencies from the joint power-law fit; column[4] the spectral index dominant at higher frequencies from the joint power-law fit; column [5] the opacity index corresponding to the high frequency power-law; column [6] the flux density predicted by the high frequency power-law at 1.8\,cm; column [7] the disk mass calculated from the flux density in column [6] assuming the opacity index $\beta'$; column [7] the disk mass calculated from the flux density in column [6] assuming an opacity index of $\beta=1$.\label{tab:beta}}
\small
\begin{tabular}{l|c|ccc|ccc}
\hline\hline
&&&&&&\multicolumn{2}{c}{$M_{\rm{disk}}$}\\
\cline{7-8}
Name & $\alpha_{\rm{AMI}}$ &  $\alpha$ & $\alpha'$ & $\beta'$ & $S_{\rm{pred,16}}$ & ${\beta=\beta'}$ & ${\beta=1}$ \\
     &                     &           &           &          &     ($\mu$Jy)      &  (M$_{\odot}$)   &  (M$_{\odot}$)   \\
\hline
IRAS04016+2610   & 0.65    &  $-1.1$   &   2.4     &   0.4    &      399           &  0.013 	 & 0.150 \\
DGTauB           & 1.79    &  $-0.1$   &   2.0     &   0.0    &      442           &  0.003	 & 0.166 \\
IRAS04248+2612   & $-$	   &  $-$      &   2.7     &   0.7    &      $<125$        &  $<0.012$   & $<0.047$\\
IRAS04302+2247   & $-$     &  $-0.1$   &   2.3     &   0.3    &       181          &  0.004      & 0.068 \\
IRAS04325+2402   & 1.49    &  $-$      &   2.1     &   0.1    &       238          &  0.002      & 0.089 \\
IRAS04361+2547   & 2.82    &  $-3.5$   &   2.0     &   0.0    &       449          &  0.003      & 0.168 \\
IRAS04368+2557   & 1.17    &  $-0.1$   &   2.3     &   0.3    &       174          &  0.004  	 & 0.065 \\
\hline
\end{tabular}
\end{table*}

\section{Discussion}
\label{sec:disc}

\subsection{Spectral Indices}

Using sub-millimetre data to derive both opacity indices and disk masses is common practice, however these values are uncertain due to both the approximately linear dependency on dust temperature, $T_{\rm{d}}$, and also the potentially significant fraction of the sub-millimetre emission that is optically thick. Perhaps more importantly, at shorter wavelengths the measured emission cannot be exclusively attributed to the disk alone and may be contaminated by the protostellar envelope. Measuring the emission at longer wavelengths reduces the effect of at least two of these issues significantly as the optically thick portion of the emission will be significantly reduced, if not negligible, by centimetre wavelengths, and additionally at frequencies below $\approx 300$\,GHz the disk component will be expected to dominate the SED in the majority of cases. 

Consequently, although the collisional growth of dust grains in the disks around young stars has been inferred from sub-millimetre observations, it is preferable to measure its properties directly from longer wavelength observations. The theoretically determined values of the opacity index for populations of these large grains ($\beta \leq 1$; Miyake \& Nakagawa 1993; Dullemond \& Dominik 2005) are consistent with those those measured from the centimetre spectrum of the objects in this sample. In the majority of cases for this sample, the contribution of additional radio emission mechanisms at 1.8\,cm has found to be sub-dominant with $\geq 70$~per~cent of the flux density attributable to thermal dust emission. IRAS~04368+2557 is an exception to this with only $20$~per~cent of the 1.8\,cm emission coming from thermal dust. This is notable as IRAS~04368+2557 is the only object in this sample which is Class~0, rather than Class~I. This lack of free--free emission in the wider sample is of particular note as such emission in young stellar objects is often attributed to the presence of a molecular outflow, with shock ionization in the outflow proposed to be the source of the ionised plasma giving rise to the free--free emission. All the objects within this sample are known to possess molecular outflows (Gramajo et~al. 2010), however as has been observed in larger samples of Class~0 \& I objects the detection rate at cm-wavelengths for Class~I sources ($\approx 20$~per~cent) is significantly lower than that of Class~0 sources ($\approx 70$~per~cent; AMI Consortium: Scaife et~al. 2011). However we note that the contribution of both thermal and non-thermal emission in young stellar objects appears to be variable (e.g. Shirley et~al. 2007) as indicated by the discrepant flux densities at radio frequencies from observations at different epochs in the literature. 

Extrapolating power-law SEDs from the sub-mm to the AMI-LA waveband requires shallow spectral indices of $\alpha = 2.0-2.7$, with an average value of $\bar{\alpha}=2.26$ corresponding to $\bar{\beta}=0.26\pm0.22$. These indices are consistent with grain growth in these objects providing a population of larger dust grains or $pebbles$. Such dust populations, with grains exceeding centimetre sizes (Draine 2005) have been theoretically demonstrated to have opacity indices approaching zero (Draine 2005; Miyake \& Nakagawa 1993; Dullemond \& Dominik 2005) and the SEDs for the sample presented in this work support these predictions.

\subsection{Disk Masses}

Using the SED fits to the radio and sub-mm data we can separate the contributions from high frequency dust emission and low frequency radio emission at 1.8\,cm. The contribution of the thermal dust emission at this wavelength is listed for each source in Table~\ref{tab:beta}. Making the assumption that all dust emission at this wavelength will be optically thin we calculate disk masses using
\begin{equation}
M_{\rm{disk}} = \frac{S_{\nu}d^2}{\kappa_{\nu}B_{\nu}(T_{\rm{d}})}.
\end{equation}

We use a dust temperature of $T_{\rm{d}}=20$\,K, the characteristic temperature found by Andrews \& Williams (2005) at 1.3\,mm. We note that increasing the value of $T_{\rm{d}}$ to that of 60\,K assumed by Terebey et~al. (1992) at 2.7\,mm would decrease our derived mass values by a factor of $\approx 3$. As well as temperature, the disk mass is a sensitive function of $\beta$ due to its linear dependence on the opacity, $\kappa_{\nu}$. Following Beckwith et~al. (1990), we assume $\kappa_{\nu} = 0.1(\nu/1000\,{\rm{GHz}})^{\beta}$\,cm$^2$\,g$^{-1}$ (a dust to gas ratio of 1:100). With $\beta=1$ this leads to a value for the opacity at 1.8\,cm (16\,GHz) of $\kappa_{16\,{\rm{GHz}}} = 0.002$\,cm$^{2}$\,g$^{-1}$.

The disk mass values derived from sub-mm data by Andrews \& Williams (2005) assume a value of $\beta=1$ when calculating the opacity. Disk mass estimates made directly from the AMI-LA flux densities under the same assumption are systematically higher than those derived from sub-mm data, see Table~\ref{tab:beta}. Drawing a comparison with the disk mass values from the best fitting SED models of Robitaille et~al. (2007) plotted in Figs.~\ref{fig:sed1},~\ref{fig:sed2} \&~\ref{fig:i04368} for each source, there is a loose agreement between the sub-mm disk mass estimates, although the scatter is large. However, the cm-wave mass estimates from the AMI-LA data are systematically higher than those derived from shorter wavelengths. One possible physical explanation for this difference is that the larger mass estimates arise as consequence of increased emission from larger dust grains at cm-wavelengths through the disk. This emission does not contribute to the sub-mm flux density, which predominantly traces smaller grains nearer to the disk surface due to optical depth effects in the disk core.

However, the disk mass estimates from the AMI-LA data using values of $\beta=\beta'$ derived from the SED fits, see Table~\ref{tab:beta}, are systematically smaller than those found in Andrews \& Williams (2005). This is a strong function of the lower values of the opacity index and therefore the correspondingly larger opacities. A clear example of this is IRAS04361+2547, where the difference in calculated disk mass is over an order of magnitude. It is unclear whether the apparent difference between the sub-mm derived mass values and those from the AMI-LA data is real. The sub-mm values use a value of $\beta=1$ which is significantly larger than that derived from SED fitting in AW05, where $\beta$ is found to have an average value of approximately zero, and so the sub-mm masses could be considered to be overestimated. Conversely the opacity index is expected to flatten towards lower frequencies and therefore the disk masses derived at cm-wavelengths should probably assume lower values of $\beta$.

Similarly to Andrews \& Williams (2005) we find that in the disk masses derived from the cm-wave flux densities assuming $\beta=1$, although larger than those inferred from sub-mm data, are still often significantly lower than those required by theory for giant planet formation (Pollack et~al. 1996; Boss 1998). Three objects are exceptions to this: IRAS04016+2610, DGTau~B and IRAS04361+2547. For each of these objects the calculated disk mass is above the limit required for planet formation of 0.13\,M$_{\odot}$ (Boss 1998). 

The gravitational instability mechanism of giant planet formation is a short lived process and most likely to occur during the early protostellar stages, at a stellar lifetime of $\approx 10^5$\,yr (Boss 1998). This lifetime is typical of Class~I objects, and given the frequency of such planets observed around more evolved stellar objects, we might consequently expect that objects such as those found in this sample would in the majority have observationally measured masses compatible with such a process. As suggested by Andrews \& Williams (2005), it is possible that an  underestimation of the disk mass from the disk SED may arise from the presence of a population of large grains which do not contribute to the emission at sub-mm to cm-wavelengths, but which acts to increase the mass of the disk. Such a situation is more likely at sub-mm than cm-wavelengths, however it is difficult to state that disk mass estimates from sub-mm flux densities are universally underestimated as there will also be a competing contribution at these wavelengths from the protostellar envelope. This contribution can only be separated out using high resolution interferometric observations. Alternatively, the core accretion mechanism of giant planet formation (e.g. Pollack et~al. 1996) requires significantly longer timescales of order $10^7$\,yr, allowing young stellar objects in the relatively early stages of their evolution to continue accreting mass onto their protostellar disks and consequently to increase their mass.

\section{Conclusions}
\label{sec:conc}

We have extended the SEDs of seven young stellar objects down to cm-wavelengths in order to constrain that part of the spectrum dominated by their circumstellar disks. The key results from this sample are summarised here:

\begin{itemize}
\item We find that for the Class~I objects in this sample the emission at 1.8\,cm is still dominated by the tail of the dust greybody spectrum, rather than alternative radio emission mechanisms, with $70-100$~per~cent of the cm-wave emission being attributable to thermal dust emission. The same is not true for IRAS~04368+2557, which is assumed to be Class~0 and has only a $20$~per~cent contribution from thermal dust at 1.8\,cm.

\item We find spectral indices consistent with a flattening of the opacity index towards longer wavelengths, with an average value $\bar{\beta}=0.26\pm0.22$, consistent with a significant population of large dust grains, and in accordance with theoretical models for the collisional growth of grains within circumstellar disks.

\item We derive disk masses directly from the thermal dust contribution to the cm-wave flux densities. Under an assumption of $\beta=1$, following Andrews \& Williams (2005) we find that these mass estimates are systematically higher than those determined from sub-mm data. Since the average value of the opacity index from SED fitting is similar at 1.8\,cm and 850\,$\mu$m, we attribute this difference in disk masses to an increased emission contribution from larger dust grains within the disk at centimetre wavelengths. Under these assumptions disk masses in excess of the lower limit required for giant planet formation (Boss 1998) are recovered in almost 50~per~cent of cases.

\end{itemize}

\section{ACKNOWLEDGEMENTS}
We thank the staff of the Lord's Bridge Observatory for their
invaluable assistance in the commissioning and operation of the
Arcminute Microkelvin Imager. The AMI-LA is supported by Cambridge
University and the STFC. MO and MS   
acknowledge the support of STFC studentships. YP acknowledges support from the Cambridge Commonwealth Trust and the
Cavendish Laboratory. AS would like to acknowledge support from Science 
Foundation Ireland under grant 07/RFP/PHYF790.

\end{document}